\documentclass[lettersize,journal]{IEEEtran}

\usepackage{xcolor} 
\usepackage{ifthen} 
\usepackage{fontawesome} 
\usepackage{listings}

\definecolor{mygreen}{HTML}{02818a}
\definecolor{nord10}{HTML}{5E81AC}
\definecolor{darkorange}{HTML}{FE6741}
\definecolor{lightred}{rgb}{1,0.9,0.9}

\newcommand{\eg}[1]{(e.g., #1)}
\newcommand{\ie}[1]{(i.e., #1)}
\newcommand{\etal}{et al.}
\newcommand{\code}[1]{{\ttfamily#1}}

\usepackage{orcidlink}
\usepackage[utf8]{inputenc}
\usepackage{xcolor,colortbl}
\usepackage{graphicx}
\usepackage{color}
\usepackage{pifont}
\usepackage{ifthen}
\usepackage{multirow}
\usepackage{booktabs}
\usepackage[inline]{enumitem}
\usepackage[ruled,linesnumbered]{algorithm2e}
\usepackage{tcolorbox}
\usepackage{booktabs}
\usepackage{tabularx}
\newcolumntype{L}{>{\raggedright\arraybackslash}X}%
\newcolumntype{R}{>{\raggedleft\arraybackslash}X}%
\newcolumntype{C}{>{\centering\arraybackslash}X}%
\usepackage{xspace}
\usepackage{fontawesome}
\usepackage{balance}
\usepackage{listings}
\usepackage{url}
\usepackage{xurl}
\usepackage{tikz}
\usetikzlibrary{trees}
\usepackage{forest}
\usepackage{float}
\usepackage{amsmath,amsfonts}
\usepackage{array}
\usepackage[caption=false,font=normalsize,labelfont=sf,textfont=sf]{subfig}
\usepackage{textcomp}
\usepackage{stfloats}
\usepackage{verbatim}
\usepackage{graphicx}
\usepackage{cleveref}
\usepackage{cite}

\lstdefinelanguage{JavaScript}{
	keywords={const, let, async, await, typeof, new, true, false, catch, function, return, null, try, catch, switch, var, if, in, while, do, else, case, break},
	ndkeywords={class, export, boolean, throw, implements, import, this},
	sensitive=false,
	comment=[l]{//},
	morecomment=[s]{/*}{*/},
	morestring=[b]',
	morestring=[b]"
}

\usetikzlibrary{shapes,positioning}
\makeatletter
\let\old@lstKV@SwitchCases\lstKV@SwitchCases
\def\lstKV@SwitchCases#1#2#3{}
\makeatother
\usepackage{lstlinebgrd}
\makeatletter
\let\lstKV@SwitchCases\old@lstKV@SwitchCases

\lst@Key{numbers}{none}{%
    \def\lst@PlaceNumber{\lst@linebgrd}%
    \lstKV@SwitchCases{#1}%
    {none:\\%
        left:\def\lst@PlaceNumber{\llap{\normalfont
                \lst@numberstyle{\thelstnumber}\kern\lst@numbersep}\lst@linebgrd}\\%
        right:\def\lst@PlaceNumber{\rlap{\normalfont
                \kern\linewidth \kern\lst@numbersep
                \lst@numberstyle{\thelstnumber}}\lst@linebgrd}%
    }{\PackageError{Listings}{Numbers #1 unknown}\@ehc}}
\makeatother

\usepackage[T1]{fontenc}

\ifx\showcomments\undefined
	\newcommand{\revised}[1]{#1}
\else
	
	\newcommand{\revised}[1]{\textcolor{blue}{#1}}
\fi

\definecolor{nord0}{HTML}{2E3440}
\definecolor{nord1}{HTML}{3B4252}
\definecolor{nord2}{HTML}{434C5E}
\definecolor{nord3}{HTML}{4C566A}
\definecolor{nord4}{HTML}{D8DEE9}
\definecolor{nord5}{HTML}{E5E9F0}
\definecolor{nord6}{HTML}{ECEFF4}
\definecolor{nord7}{HTML}{8FBCBB}
\definecolor{nord8}{HTML}{88C0D0}
\definecolor{nord9}{HTML}{81A1C1}
\definecolor{nord10}{HTML}{5E81AC}
\definecolor{nord11}{HTML}{BF616A}
\definecolor{nord12}{HTML}{D08770}
\definecolor{nord13}{HTML}{EBCB8B}
\definecolor{nord14}{HTML}{A3BE8C}
\definecolor{nord15}{HTML}{B48EAD}

\definecolor{duck}{HTML}{D89D00}

\definecolor{summarybg}{HTML}{E1EDFC}
\definecolor{findingbg}{HTML}{FFF8E6}

\newcounter{rq}

\definecolor{changesbg}{HTML}{FFF8E6}
\definecolor{changestitle}{HTML}{FF894F}

\definecolor{commentsbg}{HTML}{DDF4E7}

\newcounter{comment}[section]

\begin{document}

\title{Characterizing Tests in IoT Software: Practices, Challenges and Opportunities}

\author{%
\IEEEauthorblockN{Rufeng~Chen\orcidlink{0009-0001-2983-5699}\IEEEauthorrefmark{1}, Hengcheng~Zhu\orcidlink{0000-0002-3082-5957}\IEEEauthorrefmark{2}, Wuqi~Zhang\orcidlink{0000-0001-8039-0528}\IEEEauthorrefmark{2}, Zixu~Zhou\orcidlink{0009-0006-7536-9824}\IEEEauthorrefmark{1}, and Lili~Wei\orcidlink{0000-0002-2428-4111}\IEEEauthorrefmark{1}}\\
    \IEEEauthorblockA{\IEEEauthorrefmark{1}Department of Electrical and Computer Engineering, McGill University, Montreal, Canada}\\
    \IEEEauthorblockA{\IEEEauthorrefmark{2}The Hong Kong University of Science and Technology, Clear Water Bay, Hong Kong SAR}\\
    Emails: \IEEEauthorrefmark{1}\{rufeng.chen, zixu.zhou, lili.wei\}@mcgill.ca
    \IEEEauthorrefmark{2}\{hzhuaq, wzhangcb\}@connect.ust.hk
    \thanks{Corresponding author: Lili Wei (lili.wei@mcgill.ca).}
}

\maketitle

\begin{abstract}
	The Internet of Things (IoT) is experiencing rapid growth.
Smart devices are emerging in smart homes and industrial applications, performing mission-critical tasks. Bugs in IoT software can lead to severe consequences.
For example, a buggy smart lock can allow unauthorized access to a private property.
Testing is a primary practice to expose software bugs and ensure software quality.
However, little is known about how IoT software is tested.
To bridge this gap, we conducted the first empirical study on test cases in open-source IoT software.
Specifically, we evaluated the effectiveness of test cases in IoT software, explored the challenges inherent in testing IoT software, and analyzed the usage of mock objects.
Our results indicate that while IoT software often contains a considerable number of tests, their effectiveness remains limited.
We identified the primary challenges in testing IoT software as managing complex interactions with various external dependencies, such as other network-reliant IoT components, file systems, operating systems, and databases.
We also observed that the use of mock objects in IoT software closely aligns with our identified testing challenges.
This alignment demonstrates the potential of mocking as a solution to enhance test coverage and address the complexities of IoT software testing.

\end{abstract}

\maketitle

\section{Introduction}
\label{sec:introduction}

\IEEEPARstart{T}{he} Internet of Things (IoT) is emerging—it is estimated that the global IoT market size will grow from \$300.3B in 2021 to \$650.5B by 2026~\cite{iotforecast}.
IoT interconnects various smart devices equipped with sensors and actuators to perform mission-critical tasks that demand high reliability, security, and responsiveness in real-world environments.
For example, IoT systems can manage smart devices in a household, including light bulbs, heaters, and door locks.
Software is the key bridge that connects devices to form the Internet of Things.
Bugs in IoT software can significantly compromise the reliability and security of IoT systems and cause unprecedented consequences.
For instance, a bug in the software of a smart lock can allow unauthorized access to a private property.

Given the importance of IoT software, an increasing number of studies have been conducted to understand the challenges in developing reliable IoT software.
These studies identified interesting IoT topics~\cite{uddin2021empirical}, shared development experiences~\cite{li2022just}, reviewed testing strategies~\cite{10433067}, investigated IoT bugs and vulnerabilities~\cite{DBLP:conf/icse/Makhshari021,neshenko2019demystifying,ding2018safety}, and examined the perspective of developers in IoT platforms~\cite{10988986}.
Although these studies offer valuable perspectives on what makes IoT development and testing challenging, there remains a lack of empirical knowledge on how IoT software is tested in practice.
Little is known about the characteristics and effectiveness of existing test cases: how well they cover the code, and what factors contribute to test gaps.
These are the questions this study seeks to answer.

The heterogeneous and dynamic interdependencies of IoT systems require thorough testing to ensure reliability and security. Such systems face challenges that are rarely encountered in traditional software operating under more homogeneous and predictable conditions.
For example, unlike general software that predominantly relies on HTTP, IoT systems operate using a diverse range of communication protocols, including Zigbee, MQTT, or device-specific ones.
This diversity requires specialized testing approaches to accurately simulate and test each protocol's intricate behaviors and interactions.
In addition, IoT software must function in both digital and physical domains, leading to a vast input space and numerous execution scenarios.
Finally, the interplay between cloud, edge, and device-level code creates a multi-layered execution context, where bugs may emerge only through cross-layer interactions.
In practice, such interactions are particularly difficult to test, since IoT systems are often developed as separate components (e.g., device firmware, gateways, and cloud services) maintained in different repositories, and developers frequently test individual software parts without full access to the complete end-to-end system.
However, the unique challenges of testing in such environments remain largely underexplored by existing research, leaving a gap in enhancing IoT software testing.

To bridge this gap, we conducted the first empirical study by investigating the tests in open-source IoT software to answer the following research questions.
\\ \textbf{RQ1 (Test Effectiveness):} \textit{How effective are existing tests in IoT software? What statement coverage, branch coverage, and mutation scores can be achieved by the existing tests?}
\\ \textit{\textbf{Motivation:}} RQ1 can establish a basic understanding of the state of testing in IoT software.
By evaluating the coverage metrics, we can estimate how thoroughly IoT systems are being tested and identify specific components or functionalities where tests remain limited.
\\ \textit{\textbf{Result:}} We used static analysis on 824 IoT projects and conducted in-depth coverage and mutation analysis on 37 projects with runnable tests.
The analysis of RQ1 revealed that the tests in IoT software achieve an average statement coverage of 65.2\%, branch coverage of 53.4\%, and mutation score of 39.9\%.
Such insufficient coverage results suggest the need for a more in-depth exploration to identify the barriers to testing uncovered code (RQ2).
\\ \textbf{RQ2 (Factors Contributing to Uncovered Code):} \textit{What factors contribute to the presence of uncovered code by IoT software tests?}
\\ \textit{\textbf{Motivation:}}
RQ2 aims to discover the underlying factors that prevent certain code segments in IoT projects from being adequately tested.
Understanding these factors can guide the development of tools and methods to improve the effectiveness of IoT software testing.
\\ \textit{\textbf{Result:}} Over half of the analyzed code segments (51.4\%) remain untested primarily due to overlooked interactions with external dependencies, with untested communication scenarios representing the most significant barrier. These scenarios include network operations such as sending requests to external APIs, retrieving data from remote servers, and messaging between IoT devices.
This highlights the critical need for new testing strategies tailored to the challenges posed by the excessive external dependencies of IoT software.
\\ \textbf{RQ3 (Mocking in IoT Testing):} \textit{How are mock objects used in IoT software testing? What are their roles and what dependencies do they simulate?}
\\ \textit{\textbf{Motivation:}} IoT systems highly rely on external dependencies, and mock objects are commonly used to simulate dependencies.
This question investigates the use of mock objects in IoT software testing and explores the potential to leverage mock objects to address testing challenges for IoT software.
\\ \textit{\textbf{Result:}} The vast majority (76.04\%) of the mock objects in the test cases are simulating communication APIs such as sending and handling device interactions and network messages.
This is in line with our previous finding that the major hurdle to improving code coverage in IoT software is untested communication scenarios.
The prevalent use of mock objects emphasizes their potential application in enhancing test coverage and addressing the challenges of IoT software testing.
Meanwhile, we observe that 99.89\% of the analyzed mock objects involve defining or verifying specific dependency behaviors rather than simply acting as placeholders, making it challenging to effectively generate meaningful mock objects and enhance IoT software testing.

Collectively, these research questions outline the status quo of IoT testing practices, identify gaps and challenges, and suggest future research directions.
Our key observations suggest a promising future research direction to migrate mock objects between IoT projects by reusing the large number of readily available mock objects in open-source IoT projects.

To summarize, this paper makes the following contributions:
\begin{itemize}[leftmargin=*]
	\item We conducted the first empirical study focusing on characterizing the testing practices in open-source IoT software.
	      Our study examines how developers test IoT systems and how these practices differ from those of traditional software.
	\item We evaluated the prevalence and effectiveness of the existing tests in open-source IoT software.
	      Our analysis shows that while IoT projects frequently include test cases, these tests often fall short in effectively detecting faults, highlighting the need for improved test adequacy.
	\item We conducted an in-depth analysis to summarize the key challenges in improving the code coverage for testing in IoT software and identified the key challenge as how to properly trigger the diverse communication scenarios between IoT software and its dependencies.
	\item We analyzed the usage of mock objects in the test cases and proposed a promising future research direction to improve IoT software testing by migrating mock objects.
\end{itemize}
Our dataset, per-project metadata, and all results for reproducing the coverage and mutation analyses are publicly available in our replication package on GitHub~\cite{Github:IoTTestingAnalysis}.
\section{Background}
\label{sec:related-work-background}
\textbf{Internet of Things.}
The Internet of Things refers to a network of interconnected physical devices that collect, exchange, and process data through the internet.
These devices range from consumer products, such as smart home appliances, to industrial systems, such as manufacturing sensors. 
By integrating sensing, computation, and communication capabilities, IoT enables real-time monitoring, automation, and data-driven decision-making across diverse domains.

\textbf{IoT Software.}
IoT software refers to software across the device, edge (gateway), cloud, and application layers in IoT systems whose primary responsibility is to manage, coordinate, or encapsulate device communication and protocol-level interactions between physical IoT devices and networked services.
IoT software may overlap with embedded firmware or traditional cloud applications.
However, IoT software is distinguished by its device-facing responsibilities, such as handling protocol-specific communication, managing device identity and state, and coordinating interactions across heterogeneous physical devices and services.

\begin{figure}[t]
	\centering
	\includegraphics[width=\linewidth]{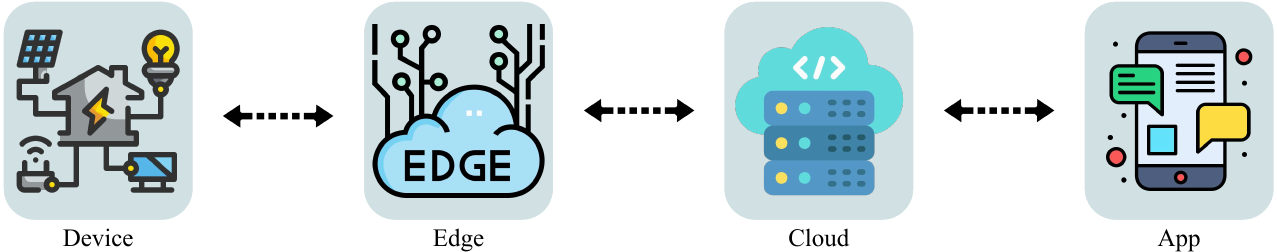}
	\caption{The Overview of IoT Layers}
	\label{fig:iot_layer}
\end{figure}
IoT software is commonly described using a layered architecture~\cite{DBLP:conf/icse/Makhshari021, 9464063} consisting of application, cloud, edge, and device layers, as illustrated in Figure~\ref{fig:iot_layer}. In the literature, these layers are sometimes referred to using alternative terminology, e.g., the edge layer is also described as the network or connectivity layer, while the cloud layer is often described as the service layer or platform layer~\cite{s24165320, 10.3389}.
Unlike traditional software, functionality in IoT systems often spans multiple layers and relies on device communication, external services, and runtime configuration.
First, IoT software must handle device-specific protocols and complex state transitions across layers, making it difficult for test suites to exercise various execution scenarios and achieve strong coverage or fault-detection effectiveness.
This directly motivates our study of test effectiveness and uncovered code in IoT projects (RQ1 and RQ2).
Second, IoT software frequently relies on external parties such as MQTT brokers, cloud APIs, gateways, or physical devices to fulfill its functionality.
As these dependencies are often unavailable during testing, developers commonly require isolation and emulation mechanisms (e.g., mocks) to enable reproducible execution.
\revised{In IoT systems, mocks are used to simulate interactions with external components, such as devices, network services, and data sources, enabling controlled and reproducible testing.}
This motivates our investigation of mocking practices in IoT software testing (RQ3).
\begin{itemize}[noitemsep, topsep=1pt, leftmargin=*]
    \item \textbf{Application Layer}: Provides user services, such as mobile apps, dashboards, and user interfaces. 
    Unlike conventional applications, IoT applications often handle heterogeneous devices, real-time monitoring, and continuous connectivity.
    \revised{In this layer, mocks are used to simulate external data sources or service responses so that application logic can be tested without relying on live systems.}

    \item \textbf{Cloud Layer}: Manages large-scale data storage, processing, and analytics. 
    It hosts APIs, machine learning models, and integration with other systems.
    IoT clouds are distinct in that they need to ingest massive streams of sensor data, ensure low-latency responses for control operations, and integrate with domain-specific protocols and devices.
    \revised{Here, mocks are used to emulate incoming data streams or external service interactions to enable scalable and repeatable testing.}

    \item \textbf{Edge Layer}: Bridges devices and cloud by using lightweight protocols (e.g., MQTT, CoAP) to filter data and enable low-latency responses under limited connectivity.
    \revised{In this layer, mocks are used to simulate device communications or network conditions to support testing of data processing and protocol handling.}

    \item \textbf{Device Layer}: Consists of hardware such as sensors, actuators, and embedded controllers that collect data and execute commands.
    IoT devices must interact with the physical world in real time.
    \revised{At this layer, mocks are used to emulate device inputs or outputs when physical hardware is unavailable or impractical to use during testing.}
\end{itemize}
Our study does not exclude any of these layers.
We analyze the testing artifacts available in open-source IoT repositories as they appear in practice.
Since IoT systems are often decomposed into separately maintained components, individual repositories may implement only a subset of the architecture.
We therefore report the layer distribution of the studied repositories in Section~III to clarify the scope of our analysis.

At the device layer, IoT testing may involve Software-in-the-Loop (SiL), Hardware-in-the-Loop (HiL), or Model-in-the-Loop (MiL) techniques.
\revised{In the open-source IoT repositories analyzed in this study, testing practices primarily adopt Software-in-the-Loop approaches, such as protocol-level simulation and virtual device responses, and we did not observe Model-in-the-Loop infrastructure.}
\revised{Accordingly, our empirical results primarily reflect software-level testing.}

Prior secondary studies have examined software testing challenges at different IoT layers.
Garousi \etal{} provide a comprehensive survey of embedded software testing techniques, including device-level validation and simulation-based testing~\cite{10.1109/MS.2018.2801541}.
More recent surveys analyze testing challenges in distributed IoT systems spanning device, edge, and cloud layers, highlighting issues such as heterogeneity, limited observability, and complex dependencies~\cite{10433067,10.1145/3625094}.
These studies collectively indicate that IoT software testing faces stronger environmental constraints and dependency complexity than general-purpose software systems.

Unlike traditional software systems, IoT software often interacts with various communication protocols, operates across heterogeneous hardware platforms, and adapts to dynamic and distributed environments.
These characteristics pose unique testing challenges, particularly in designing effective test inputs. 
\revised{Test cases must follow communication protocols across devices, networks, and cloud services, while reflecting realistic environmental conditions (e.g., temperature, motion, or location) and continuous interaction with the physical world.}

\section{Data Collection}\label{sec:dataset}
\begin{figure*}[t]
	\centering
	\includegraphics[width=0.9\linewidth]{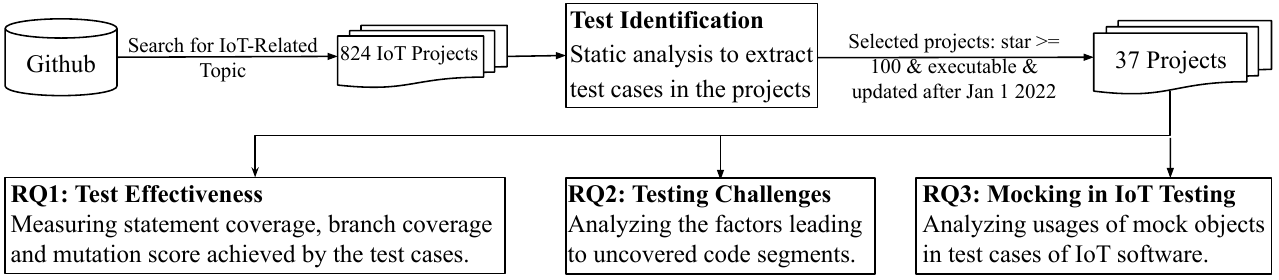}
	\caption{The Overview of Our Empirical Study}
	\label{fig:overview}
\end{figure*}

Figure~\ref{fig:overview} shows an overview of the workflow of our empirical study.
We started by collecting a dataset of 824 GitHub projects related to IoT.
Makhshari and Mesbah~\cite{DBLP:conf/icse/Makhshari021} also collected a dataset of IoT software. Although their dataset contains 91 IoT projects, many of these projects are not suitable for dynamic test execution: 29 projects had been abandoned for at least four years, and 65 do not build or run.
Since answering our research questions requires executing the tests of IoT software, we collected this new dataset with more up-to-date IoT software repositories, which are more likely to build and run successfully~\cite{Sul_r_2016}.
Using static analysis, we extracted test cases from these 824 projects and summarized their distribution.
Among the 317 projects that contained test cases, 88 projects with more recent updates and higher popularity were further considered for detailed analysis.
Although only 37 of these projects could be successfully executed, \revised{they are representative of the larger set, accounting for 12.5\% of the total lines of code and 20.4\% of the test cases, and covering all four IoT layers.}
These 37 projects are representative of our dataset, given their substantial number of test cases and their coverage of all IoT layers.
These account for 12.5\% of total LOC and 20.4\% of test cases observed across the 317 projects.
At the 95\% confidence level, this sample size corresponds to a margin of error of approximately $\pm$5\% when calculated with respect to the total number of test cases, indicating that our dataset provides a representative basis for analysis.
We compared the 37 selected projects with the full set of 317 IoT repositories at an aggregate level.
As shown in Table~\ref{tab:dataset_comparison}, both groups exhibit similar project sizes, measured by the average number of source files per project.
The selected projects contain more test cases and substantially lower P/T \revised{(Production-to-Test)} ratios on average, reflecting the presence of active and executable test suites required to obtain meaningful coverage and mutation results.
Projects with sparse or inactive tests tend to yield trivial or uninformative dynamic analysis outcomes.
Such outcomes are uninformative because they primarily reflect missing or broken test activity rather than the challenges of achieving effective testing in IoT software.
Therefore, we exclude them to avoid drawing misleading conclusions.
We note that this selection criterion may bias the reported coverage and mutation scores toward better-tested IoT repositories.
Accordingly, the effectiveness metrics reported in this study should be interpreted as characterizing test-active and runnable IoT projects, rather than the entire population of IoT repositories.
In addition, the selected projects collectively cover all IoT layers with comparable distributions, preserving architectural diversity.
To validate this observation, we performed two-proportion z-tests for each layer and found no statistically significant differences between the selected subset and the full dataset.
In line with our motivation, we treat each repository as an independently developed IoT software component, reflecting the realistic setting where developers often have limited access to the full IoT system.
Thus, the analyzed projects may implement only one or a few IoT layers, while their tests still need to account for cross-layer dependencies through communication, configuration, and external services.
\revised{Specifically, among the 317 analyzed repositories, 7 projects implement all four layers, while 88 projects implement only a single layer, with the remaining projects covering multiple layers.
The complete mapping between projects and their corresponding layers is available in our online repository.}
In this section, we detail the data collection process and describe our approach for each research question separately before discussing their results.

\begin{table}[H]
\centering
\caption{\revised{Comparison Between Selected and Full-Set IoT Repositories}}
\label{tab:dataset_comparison}
\small
\resizebox{\columnwidth}{!}{
\begin{tabular}{lrr}
\toprule

\textbf{\revised{Metric}} & \textbf{\revised{37 Selected Projects}} & \textbf{\revised{317 IoT Repositories}} \\
\midrule
\revised{Number of repositories} & \revised{37} & \revised{317} \\
\revised{Average source files per project} & \revised{499.4} & \revised{518.7} \\
\revised{Average test cases per project} & \revised{813.3} & \revised{466.0} \\
\revised{Average P/T ratio} & \revised{23.0} & \revised{356.9} \\
\revised{Device-layer} & \revised{13 (35.1\%)} & \revised{112 (35.3\%)} \\
\revised{Edge-layer} & \revised{12 (32.4\%)} & \revised{85 (26.8\%)} \\
\revised{Cloud-layer} & \revised{20 (54.1\%)} & \revised{199 (62.8\%)} \\
\revised{Application-layer} & \revised{8 (21.6\%)} & \revised{90 (28.4\%)} \\

\bottomrule
\end{tabular}
}
\revised{\footnotesize \textit{Note:} P/T ratio denotes the Production-to-Test ratio.}
\end{table}

\subsection{Project Collection}
To determine the selection criteria for our IoT project dataset, we initially reviewed the dataset collected by Makhshari and Mesbah~\cite{DBLP:conf/icse/Makhshari021}, which includes 91 IoT projects from GitHub.
We analyzed their primary languages, defined as the language with the most files in each repository.
JavaScript, Python, Java, and C emerged as the top languages, accounting for 78.9\% of the projects.
Based on this insight, we focused on JavaScript, Python, and Java.
C was excluded due to the lack of standardized testing frameworks, which prevents automatic test case identification for further analysis.

We adopted a similar approach to Makhshari and Mesbah~\cite{DBLP:conf/icse/Makhshari021} in collecting IoT projects by using GitHub topic tags.
We sourced our dataset from GitHub, adhering to three selection criteria:
(1) The primary language of the project is either JavaScript, Python, or Java. (2) The projects should be tagged with at least one of the following keywords: IoT, Internet of Things, IoT-platform, IoT-device, IoT-Application. These tags were the top five most popular topics on GitHub containing ``IoT'' at the time of the experiment. (3) The project must have more than 10 stars to exclude toy projects.
We collected a dataset comprising 824 open-source GitHub projects. The dataset is characterized by a diverse representation of primary languages: 371 Python projects (45\%), 182 Java projects (22\%), and 271 JavaScript projects (33\%).
The dataset includes renowned smart home platforms such as Home Assistant (58.7k stars)~\cite{Github:home-assistant/core} and ThingsBoard (13.2k stars)~\cite{Github:thingsboard/thingsboard}, protocol libraries like paho.mqtt.python (1.8k stars)~\cite{Github:paho.mqtt.python}, backend SDKs such as kuzzle (1.2k stars)~\cite{Github:kuzzle}, and command-line interfaces like keylime (300 stars)~\cite{Github:keylime}.
This variety ensures that the dataset reflects the multifaceted nature of IoT projects.
Our study focuses on production-level IoT software repositories in which testing is part of normal development practice.
Repositories primarily designed for testing frameworks or benchmarking are excluded, as they do not reflect how testing is integrated into real IoT software projects.
We acknowledge such repositories as a complementary population and identify their analysis as an important direction for future work.

\subsection{Dataset Characterization}
The analyzed repositories are primarily software components rather than complete IoT systems.
We do not distinguish whether a repository represents an entire IoT system.
Instead, each repository is characterized based on the IoT layers it implements (application, cloud, edge, and device).
This characterization enables us to analyze the dataset’s architectural distribution and include projects from different layers.

\revised{
To make this distribution explicit, Table~\ref{tab:dataset_comparison} reports the layer coverage of both the 37 selected runnable projects and the full set of 317 IoT repositories.
The results show that the selected projects preserve similar architectural diversity.
Device-layer projects account for 35.1\% of the selected dataset versus 35.3\% in the full pool, while cloud-layer projects represent 54.1\% versus 62.8\%.
Application-layer components are also present (21.6\% of the selected projects).}

To further clarify what kinds of IoT software artifacts are included in our dataset, we provide examples from each IoT layer observed in our dataset. 
At the \textbf{device layer}, repositories often implement device-facing communication logic that directly interacts with sensors, actuators, or embedded protocols.
For example, \texttt{i2c-bus} provides low-level access to hardware devices over the I2C bus. 
At the \textbf{edge layer}, repositories typically provide gateway or protocol-mediation components that bridge local devices with external services.
For instance, \texttt{iotedgedev} supports IoT Edge deployment workflows.
At the \textbf{cloud layer}, many repositories implement backend services or coordination platforms that manage fleets of connected devices through IoT protocols.
A representative example is \texttt{azure-iot-sdk-python}, which provides device--cloud communication support and end-to-end messaging workflows.
Finally, at the \textbf{application layer}, some repositories include user-facing automation and management logic for configuring IoT devices and workflows.
For example, \texttt{home-assistant-cli} provides command-line interfaces for IoT administration.

\subsection{Test Identification}
\label{sec:rq1}

\subsubsection{Approach}

Based on the collected dataset, we first analyzed the prevalence of test cases in IoT software via static analysis.
We adopted a two-stage approach to identify the test code in our subjects.
In the first stage, we identified code patterns for common testing frameworks in our subjects.
In the second stage, we designed automated techniques to identify test code in these projects based on the code patterns.

\textbf{Stage I: Code Pattern Derivation.}
In the first stage, we aimed to derive code patterns of commonly used testing frameworks.
For each project, we inspected the source files and directory structures to check if they contained any test cases.
We then identified the testing frameworks by examining the test code and build configurations.
To derive the patterns, we analyzed test cases and relevant documentation, identifying common configurations and method annotations that define test cases.
The resulting code patterns are as follows:

\begin{itemize}[leftmargin=*]
	\item \textsc{JavaScript}:
	      \textsc{Jest}~\cite{jest}, \textsc{MochaJS}~\cite{mochajs}, and \textsc{Jasmine}~\cite{jasmin} are the commonly used frameworks in our \textsc{JavaScript} subjects.
	      We identified each function closure passed to the testing API \code{test()} or \code{it()} as a test case.
	      Each of these closures is recognized as an individual test case.
	\item \textsc{Python}:
	      \textsc{PyTest}~\cite{pytest} and \textsc{UnitTest}~\cite{unittest} are the two commonly used testing frameworks.
	      We identified \textsc{PyTest} test cases by matching the functions whose names start with \code{test} in the source files named with the prefix \code{test\_} or postfix \code{\_test}.
	      We identified test cases using the \textsc{UnitTest} framework by detecting methods within classes that inherit \code{unittest.TestCase}.
	\item \textsc{Java}:
	      \textsc{JUnit}~\cite{junit} is the de facto testing framework for \textsc{Java} projects.
	      We identified the methods with annotation \code{@Test} as \textsc{JUnit} test cases.

\end{itemize}

\begin{table}[]
	\centering
	\caption{Distribution of Test Cases by Primary Languages}
	\label{tab:test_cases_with_lan}
	\resizebox{\columnwidth}{!}{%
		\begin{tabular}{@{}lrrrrr@{}}
			\toprule
			\multirow{2}{*}{\textbf{\# Projects}} & \multicolumn{3}{c}{\textbf{\# Test Case}} & \multicolumn{2}{c}{\textbf{P/T Ratio}}                                                     \\ \cmidrule(l){2-6}
			                                      & \textbf{Total}                            & \textbf{Mean}                          & \textbf{Median} & \textbf{Mean} & \textbf{Median} \\ \midrule
			317                                   & 147,736                                   & 466.0                                  & 34              & 356.9         & 8.97            \\ \bottomrule
		\end{tabular}%
	}
\end{table}

\textbf{Stage II: Automated Test Identification.}
Using the derived code patterns, we implemented an automated tool to identify test cases by matching common testing frameworks patterns.
The tool first locates candidate test files, those named with “test” or placed in “test” directories.
Then it performs pattern matching on the AST of the source file.
For example, in Java projects using \textsc{JUnit}, our tool identifies methods annotated with \code{@Test}.
Details of all patterns used by our tool are available on our project's website~\cite{Github:IoTTestingAnalysis}.
In each project, we analyzed the production and test code written in its primary language.
During the analysis, we ignored the directories that are commonly used to store external libraries \eg{\code{node\_modules}, \code{platform}, and \code{third-party}}.

\subsubsection{Results}

Among the 824 projects analyzed, 317 (38.5\%) contain tests.
This proportion notably exceeds what is observed within the broader spectrum of GitHub projects, where only 17.2\% are reported to include tests~\cite{7962388}.
Furthermore, this rate is substantially higher than the 14.2\% within Android applications as identified in previous research~\cite{7102609}.
Such findings highlight the pronounced emphasis on testing within IoT software projects, surpassing general software development trends and specific domains like Android apps.
Table~\ref{tab:test_cases_with_lan} details the test case distribution across these 317 projects, with an average of 466 test cases per project and a median of 34.

The absolute number of test cases may not reflect the actual test abundance since different projects can have different scales.
To account for this variation,
we introduce the Production-to-Test (P/T) ratio, defined as the ratio of lines of production code to lines of test code.
A smaller P/T ratio corresponds to a larger proportion of test code, indicating that the underlying project is more extensively tested.

\[
	\textrm{P/T ratio} = \frac{
		\textrm{Lines of Production Code}%
	}{
		\textrm{Lines of Test Code}
	}
\]

\begin{figure}[t]
	\centering
	\includegraphics[width=\linewidth]{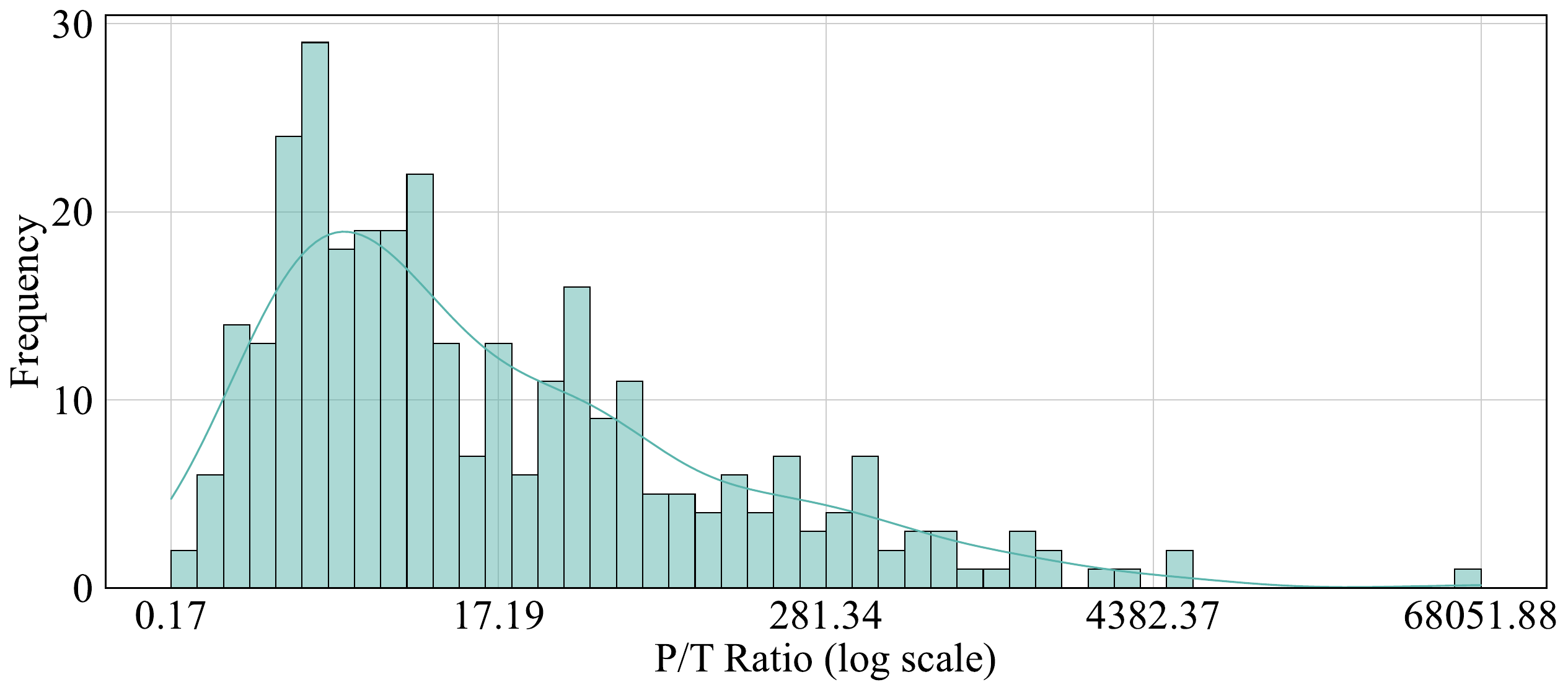}
	\caption{Distribution of P/T Ratio of IoT Projects
	}
	\label{fig:ratio}
\end{figure}

Figure~\ref{fig:ratio} illustrates the distribution of the P/T ratio.
In the context of projects with commendable code coverage, a P/T ratio lower than 1 is typically indicative of robust testing practices~\cite{khorikov2020unit}.
However, among our examined IoT projects, only 13 (4.1\%) exhibited a P/T ratio below 1, while a substantial portion (124 out of 317, or 39.1\%) have a P/T ratio exceeding 17.19, indicating that many IoT projects include very limited test code.
The project with the highest P/T ratio, Istsos2~\cite{istsos2}, an IoT server encompassing both a backend server and a GUI application, registered a P/T ratio of 68051.88, highlighting a critical lack of testing relative to its production code.
These results suggest that despite the presence of test cases, testing in many IoT projects remains insufficient compared to code size.


\subsection{Test Granularity}
The collected test cases can have varying granularities as unit tests, integration tests, or system tests.
To gain insight into this, we performed a closed coding analysis following the deductive approach by Pecorelli \etal{}~\cite{10186855}, classifying 382 Python, 383 Java, and 379 JavaScript tests (95\% CL, 5\% ME).
This classification was based on criteria such as the tested components, dependency isolation, and documentation.
Our analysis identified 779 (68.1\%) as unit tests, 348 (30.4\%) as integration tests, and 17 (1.5\%) as system tests, highlighting the predominance of unit and integration testing in the analyzed IoT projects.
Therefore, the insights derived from our subsequent analysis mainly reflect testing practices for unit and integration levels.

\section{RQ1: Test Effectiveness}
\label{sec:rq2}
\subsection{Approach}

RQ1 aims to investigate the effectiveness of the test cases in IoT software.
To answer this research question,
we evaluated the statement coverage, the branch coverage, and the mutation scores that can be achieved by the existing test cases of the selected IoT projects.
Our approach unfolds as follows:

Executing test cases of the collected projects requires tremendous manual effort to properly set up the execution environment, since some of the projects may not contain clear instructions to set up the runtime environment or the instructions are outdated.
As a result, it is impractical to run the test cases of all 317 projects.
To filter out very inactive projects while maintaining representativeness, we applied two basic selection criteria:
(1) repositories must have at least 100 GitHub stars to ensure minimal community adoption, and
(2) repositories must have at least one commit after January~1, 2022 to exclude projects with no recent development activity.
We adopted these criteria because long-term inactive projects often contain outdated dependencies or unmaintained test setups, making dynamic execution unreliable and potentially leading to uninformative coverage and mutation outcomes.

To further validate that the selected projects remain actively maintained beyond this cutoff, we examined their commit recency and commit frequency.
Only one project in the final dataset has no commits after the cutoff date.
Overall, 68\% of the selected projects were updated within 2025, and the selected projects exhibit an average commit frequency of 13 commits per month after the cutoff date (median: 2), indicating sustained maintenance beyond the cutoff. 

These filters yielded 88 projects for further analysis.
We then attempted to measure their code coverage, following each repository’s build and test instructions when available to replicate their testing setup.
Nevertheless, some projects could not be built due to obsolete dependencies, specific hardware requirements, and incomplete documentation.
In addition, the tests of some successfully executed projects finished with errors.
To ensure the integrity of our collected test coverage, we included only those projects whose tests were executed without errors.
This exclusion was essential to eliminate potential biases arising from incorrect test configurations or environmental discrepancies.
In the end, we successfully executed the tests in 37 projects (42\%).
To collect code coverage and mutation scores, we used established automated tools.
For Python projects, we used \textsc{mutmut} to perform mutation testing with all default mutation operators enabled, and \textsc{Coverage.py} to collect statement and branch coverage.
For Java projects, we used \textsc{PIT} with its standard mutation operators and \textsc{JaCoCo} for coverage measurement.
For JavaScript projects, we employed \textsc{Stryker} with all built-in mutators enabled and \textsc{nyc} for coverage collection.
All tools provide automated instrumentation and command-line execution, enabling consistent measurement across projects and facilitating replication of our experimental setup. 

\begin{figure}[t]
	\centering
	\includegraphics[width=\linewidth]{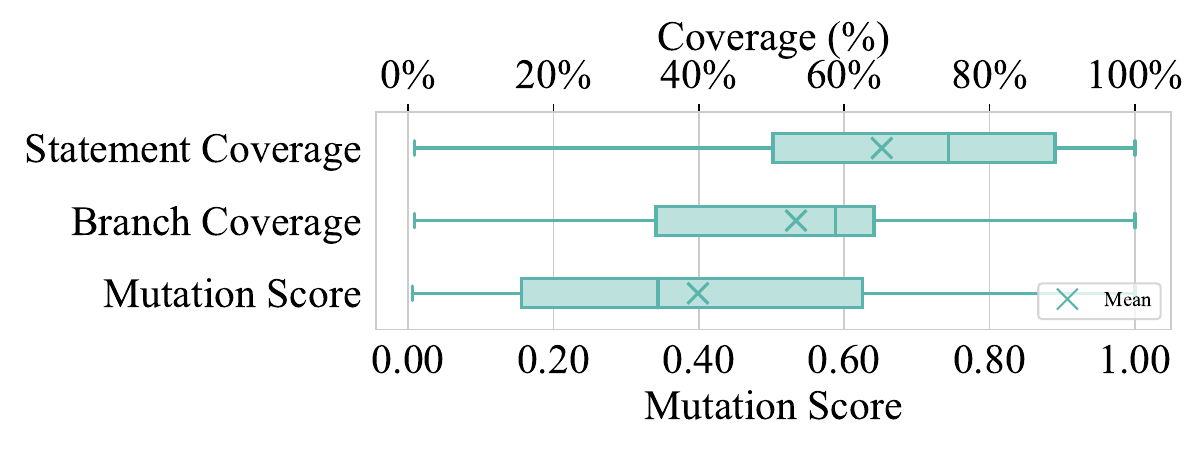}
	\caption{Boxplots showing Statement Coverage, Branch Coverage, and Mutation Score across IoT projects. Coverage is based on 37 projects; mutation scores on 28 with successful analysis.}
	\label{fig:coverage_boxplot}
\end{figure}

\subsection{Results}

Figure~\ref{fig:coverage_boxplot} shows the distribution of statement and branch coverage across analyzed projects.
The average statement and branch coverage were 65.2\% and 53.4\%, respectively.
These metrics characterize testing effectiveness among runnable and test-active IoT projects, rather than the entire population of IoT repositories.
These averages are higher than those reported in previous studies on Maven projects, which showed average statement coverages of 37.8\% and 42.0\%~\cite{8031982,7091313}, suggesting relatively stronger testing effort in the runnable and test-active IoT projects we analyzed.
However, most projects still lack completeness: 54.1\% have statement coverage below 75\%, and 75.7\% have branch coverage below 75\%.
We also observe significant disparities between projects.
The project vandium-node~\cite{Github:vandium-node} had a balanced P/T ratio of 1.24 and achieved 100\% coverage.
In contrast, the lowest statement and branch coverage were both 0.9\% for smart-socket~\cite{Github:smart-socket} and UsbSerial~\cite{Github:UsbSerial}, with high P/T ratios of 347.9 and 19.81.

Despite extensive efforts, mutation analysis failed for 9 of the 37 projects due to tool limitations, including crashes caused by bugs in the tools and compatibility issues such as conflicts with the project's build environment.
Figure~\ref{fig:coverage_boxplot} shows that test effectiveness remains limited in many analyzed IoT projects, with an average mutation score of 39.9\% and a median of 34.4\%.
This suggests that many test cases fail to effectively detect faults introduced through mutation, highlighting gaps in test effectiveness.
This indicates limited fault-detection ability in many test cases—18 out of 28 projects (64.3\%) scored below 50\%.
The highest score, 100\%, was achieved by Adafruit\_IO\_Python~\cite{Github:Adafruit}, which also had over 90\% statement coverage.
In contrast, smart-socket~\cite{Github:smart-socket} scored just 2.8\%, with only 5 test cases and 0.9\% statement coverage.

These observations suggest that substantial room for improvement remains in the testing practices of the runnable and actively tested IoT projects we analyzed, motivating our investigation in RQ2 to uncover the underlying challenges.
Understanding these challenges is essential for advancing testing methodologies.



\begin{tcolorbox}[left=3pt,right=3pt,top=1pt,bottom=1pt]
	\textbf{Answer to RQ1:}
	\textit{
		Analysis reveals average statement coverage of 65.2\%, branch coverage of 53.4\%, and mutation score of 39.9\% among runnable and test-active IoT projects.
        For context, prior studies on Maven-based Java projects reported average statement coverages of 37.8\% and 42.0\%, while empirical studies on GitHub open-source systems reported average mutation scores around 40\%.
        Compared with these baselines, IoT projects exhibit substantially higher coverage but similar mutation effectiveness, motivating further analysis in RQ2.
	}
\end{tcolorbox}

\section{RQ2: Factors Contributing to Uncovered Code}
\label{sec:rq3}
\subsection{Approach}\label{ssec:rq3approach}
The goal of RQ2 is to analyze the factors that lead to uncovered code within IoT software testing, and thereby to pinpoint the challenges in testing for IoT software.
To achieve this, we adopted an open-coding strategy to analyze the uncovered code segments identified in RQ1.

For our manual analysis, we included the projects that are the top 75\% in terms of both the highest statement coverage and the lowest P/T ratio.
We adopted such a selection criterion to exclude projects with overly low coverage or excessively high P/T ratios.
Projects with low statement coverage have extensive portions of untested code.
Similarly, a high P/T Ratio suggests that a project where the volume of production code far outweighs the extent of test code.
Limited efforts were put into testing in these projects.
Code segments in such projects may not be covered simply because the developers did not write a test for them.
Analyzing such projects cannot provide meaningful insights for this RQ.
In contrast, the selected projects with higher code coverage and lower P/T ratio are more likely to contain well-maintained and comprehensive test cases.
Inspecting the factors leading to uncovered code in these projects is more likely to provide useful insights into testing IoT software.
This selection resulted in 11 Python projects, seven JavaScript projects, and seven Java projects.
Table~\ref{tab:project_flow} summarizes the number of projects remaining at each filtering step for each language, from the initial pool to the final selected dataset.

\begin{table}[H]
	\centering
	\caption{Project Selection Flow by Language.}
	\label{tab:project_flow}
	\small
	\resizebox{\columnwidth}{!}{
		\begin{tabular}{lrrrrr}
			\toprule
			\textbf{Language}              &
			\textbf{Initial}               &
			\textbf{Have Tests}            &
			\textbf{Star \& Active Filter} &
			\textbf{Runnable}              &
			\textbf{Coverage \& P/T Filter}                           \\
			\midrule
			Python                         & 371 & 122 & 34 & 16 & 11 \\
			JavaScript                     & 271 & 80  & 13 & 7  & 7  \\
			Java                           & 182 & 115 & 41 & 14 & 7  \\
			\bottomrule
		\end{tabular}
	}
\end{table}

Subsequently, we sampled the uncovered code segments with a 95\% confidence level and a 5\% margin of error.
Each entry represents a contiguous block of uncovered code, rather than isolated lines, allowing us to analyze meaningful untested segments.
Due to differences in uncovered code volume across languages, we applied language-specific sampling: Python — 202 from 1,103 entries; JavaScript — all 131 entries analyzed; Java — 383 from 89,393 entries.

Our open coding procedure involved two authors independently categorizing uncovered code segments based on source code and corresponding test reports.
An initial round of independent analysis was followed by discussion to define a hierarchical taxonomy and establish consistent categorization criteria.
The authors then applied these criteria to classify the entire dataset, holding iterative discussions to address discrepancies and refine the taxonomy as needed.
A third author participated in resolving conflicts.
We achieved a Cohen's Kappa score of 0.87, indicating a high level of agreement.

Although this research question aims to identify factors contributing to the existence of uncovered code segments in IoT software testing, we encountered a considerable number of code segments where no obvious testing challenges could be identified.
For example, we found code segments where entire functions remained untested, yet the function arguments were straightforward to set up.
To systematically address these ambiguous cases, we adapted our classification strategy to focus on the intrinsic functionalities of uncovered code.
This approach helps reveal common, critical functionalities that are vital yet overlooked by developers in testing.

\subsection{Results}

To understand the factors leading to uncovered code segments, we derived a hierarchical taxonomy.
All the code segments are first categorized into three primary categories as shown in Table~\ref{tab:cate1}.
The first primary category (Uncovered Code Segments with Identified Hurdles) contains 368 code segments where we successfully observed the factors impeding their coverage.
We identified the subcategories as the impeding factors.
The second primary category (Uncovered Code Segments with No Identified Hurdles) contains 241 code segments without identifiable coverage hurdles.
To gain a clearer understanding of these segments, we analyzed their functionalities.
The third primary category (Unreachable Code Segments) contains 107 cases where the code segments cannot be covered, such as dead code or unimplemented features.
Below, we introduce these primary categories and subsequently delve into their respective subcategories to summarize the key challenges in achieving better test coverage for IoT software.

\begin{table}[t]
	\centering
	\caption{Taxonomy of Uncovered Code Segments}
	\label{tab:cate1}

	\begin{tabularx}{\linewidth}{@{}Xr@{}}
		\toprule
		\textbf{Subcategory}                                         & \textbf{Count} \\
		\midrule
		\midrule
		\textbf{Uncovered Code Segments with Identified Hurdles}     & 368            \\
		\midrule
		\hspace{1em}Untested Communication Scenarios                 & 199            \\
		\hspace{1em}Misaligned File System Dependency                & 57             \\
		\hspace{1em}Improper System Configuration                    & 55             \\
		\hspace{1em}Insufficient Database Setup                      & 52             \\
		\hspace{1em}Untested IoT-Event Handlers                      & 5              \\
		\midrule
		\midrule
		\textbf{Uncovered Code Segments with No Discernible Hurdles} & 241            \\
		\midrule
		\hspace{1em}CLI Usage                                        & 48             \\
		\hspace{1em}Illegal Object Handling                          & 43             \\
		\hspace{1em}Arid Code                                        & 33             \\
		\hspace{1em}Overridden Standard APIs                         & 21             \\
		\hspace{1em}Data Structure Processing                        & 15             \\
		\hspace{1em}Inherited and Overridden Functions               & 8              \\
		\hspace{1em}Others                                           & 73             \\
		\midrule
		\midrule
		\textbf{Unreachable Code Segments }                          & 107            \\
		\midrule
		\hspace{1em}Test Suite Anomalies                             & 64             \\
		\hspace{1em}Pending Implementations                          & 26             \\
		\hspace{1em}Dead Code                                        & 17             \\
		\bottomrule
	\end{tabularx}
\end{table}

\subsubsection{Category 1: Uncovered Code Segments with Identified Hurdles}

This category contains 368 code segments (51.4\%) where we identified factors impeding their coverage.
We observe that the majority of them are not covered due to the various constraints on dependencies and environments.
These constraints can arise from interactions with web services, database operations, file system manipulations, network connectivity issues, etc.
We further categorize the factors into subcategories and Table~\ref{tab:cate1} shows their distribution.
We detail the definition of each subcategory as follows:

\textbf{Untested Communication Scenarios (199 code segments):}
This subcategory encompasses code segments that require communication with various IoT components or web services, often executed through network requests using various protocols (e.g. HTTP, MQTT~\cite{MQTT}, and LWM2M~\cite{LWM2M}).
It involves operations such as sending requests to external APIs, fetching data from remote servers, or handling responses from web services and device interactions.
Listing~\ref{lst:communication} shows an example in this category from Adafruit IO Python~\cite{Github:Adafruit}.
The function subscribes to MQTT-based weather forecast topics provided by Adafruit IO, requiring both a valid weather ID and forecast type (Line 1).
Line 2 checks whether the provided forecast type is valid.
The entire function was left untested in the project.
Covering this function requires constructing input strings \ie{\texttt{weather\_id} and \texttt{forecast\_type}} that match the predefined MQTT string format.
Moreover, it also requires mocking the MQTT client and verifying that it performs correct subscriptions for valid inputs.

\definecolor{uncovered}{HTML}{FCF3D5}
\begin{figure}[t]
	\fontsize{7}{7}
	\begin{lstlisting}[language=Python, caption={Example of Communication Scenarios (highlighted lines indicate uncovered code, this also applies to the other listings)}, label=lst:communication ,gobble=8, numbers=left, ,
    		linebackgroundcolor = {\ifnum \value{lstnumber} > 1 \ifnum \value{lstnumber} < 9 \color{uncovered} \fi \fi},
    ]
        def subscribe_weather(self, weather_id, forecast_type):
          if forecast_type in forecast_types:
            self._client.subscribe(
              '{0}/integration/weather/{1}/{2}'.format(
                self._username, weather_id, forecast_type
            ))
          else:
            raise TypeError("Invalid Forecast Type Specified.")
	\end{lstlisting}
\end{figure}

\begin{figure}[t]
	\centering
	\begin{minipage}{0.48\textwidth}
		\begin{lstlisting}[language=Python, caption={Example of Misaligned File System Dependency}, label=lst:file, gobble=8, numbers=left,
    		linebackgroundcolor = {\ifnum \value{lstnumber} > 4 \ifnum \value{lstnumber} < 6 \color{uncovered} \fi \fi},
        ]
        self.json = json.loads(
          Utility.get_file_contents(path, expandvars=expand_vars)
        )
        if "modulesContent" in self.json:
          return self.json["modulesContent"]
        \end{lstlisting}
	\end{minipage}
	\hfill
	\begin{minipage}{0.48\textwidth}
		\begin{lstlisting}[language=Python, caption={Example of Improper System Configuration}, label=lst:env, gobble=8, numbers=left,
    		linebackgroundcolor = {\ifnum \value{lstnumber} > 1 \ifnum \value{lstnumber} < 3 \color{uncovered} \fi \fi},
        ]
        if "CONNECTION_STRING" in os.environ:
          # Uncovered block...
        \end{lstlisting}

	\end{minipage}
\end{figure}

\begin{figure}[t]
	\centering
	\begin{minipage}{0.48\textwidth}
		\begin{lstlisting}[language=Java, caption={Example of Insufficient Database Setup Interaction}, label=lst:databaseInteraction, gobble=8, numbers=left,
    		linebackgroundcolor = {\ifnum \value{lstnumber} > 4 \ifnum \value{lstnumber} <6 \color{uncovered} \fi \fi},
        ]
        try {
          writeBatch.put(..., rocksKey(row), ...);
          return null;
        } catch (RocksDBException e) {
          throw new StorageException(...);
        }
        \end{lstlisting}
	\end{minipage}
	\hfill
	\begin{minipage}{0.48\textwidth}
		\begin{lstlisting}[language=Python, caption={Example of Event-Driven Scenarios}, label=lst:event, gobble=8,numbers=left,
    		linebackgroundcolor = {\ifnum \value{lstnumber} > 1 \ifnum \value{lstnumber} <3 \color{uncovered} \fi \fi},
        ]
        if event == SUN_EVENT_SUNRISE:
          brightness = lerp(...)
        \end{lstlisting}
	\end{minipage}
\end{figure}

\textbf{Misaligned File System Dependency (57 code segments):}
This subcategory includes code segments involved in file processing, such as I/O operations and file system interactions.
Testing these segments is challenging due to the need to simulate file system behavior or ensure specific file contents and formats.
Listing~\ref{lst:file} shows an example code snippet in Iotedgedev~\cite{iotedgedev} where Lines 4–5 are uncovered.
Covering them requires a test case that places a JSON file containing the key \code{"modulesContent"} in the expected directory.
This type of code requires careful consideration in testing to account for various file system states, such as the existence, accessibility, and correctness of specific files.

\textbf{Improper System Configuration (55 code segments):}
This category includes code segments that rely on the underlying operating system, including dependencies on environment variables, platform-specific characteristics, and system calls.
These dependencies complicate testing, as they require the test environment to replicate specific system conditions or permissions.
Listing~\ref{lst:env} shows an example code snippet in Azure-IoT-SDK-Python~\cite{azure-iot-sdk-python}, where Line~2 checks if \texttt{CONNECTION\_STRING} exists in the environment variables of the system.
This environment variable is necessary for remote connections for this project.
Testing such code segments requires the replication of specific environmental settings.

\textbf{Insufficient Database Setup (52 code segments):}
This subcategory includes code segments that interact with databases.
Testing them is challenging due to the need to simulate connections, control database states, and validate both normal and exceptional behavior.
Listing~\ref{lst:databaseInteraction} shows an example from Ignite~3~\cite{ignite-3}, where Line 5 is uncovered. Lines 2–3 perform a database insertion via RocksDB, while Line 5 handles a possible \code{RocksDBException}.
Covering this line requires configuring RocksDB with conditions that trigger \code{RocksDBException}.
Effective testing must account for both successful operations and error handling scenarios.

\textbf{Untested IoT-Event Handlers (5 code segments):}
This category includes code segments triggered by hardware-related or environmental stimuli events, such as temperature changes, motion detection, light levels, or other sensor-based inputs.
Such scenarios are crucial for IoT software that reacts to environmental changes or user interactions.
Testing these segments is challenging due to the need to simulate various real-world conditions.
Listing~\ref{lst:event} shows an example code snippet in Adaptive Lighting~\cite{Github:adaptive-lighting} where Line~2, which adjusts brightness based on a sunrise event, remains untested.

\subsubsection{Category 2: Uncovered Code Segments with No Discernible Hurdles}
This category contains code segments where we did not observe any challenges in writing test cases to cover them.
These code segments are typically entire functions that are uncovered and arguments of the functions are straightforward to create.
It is unclear why these code segments are not covered.
As a result, we focused on categorizing the functionalities of these code segments.
241 out of the 716 (33.7\%) analyzed code segments fall into this category.

As shown in Table~\ref{tab:cate1}, the functionalities of code segments in this category fall into: (1) \textit{CLI Usage (48 code segments):}
Code segments that execute or interact with the Command Line Interface (CLI).
(2) \textit{Illegal Object Handling (43 code segments):}
Code segments that deal with illegal or unexpected object states, including handling null references, invalid object configurations, or exceptions arising from improper use of objects.
The illegal objects of these category are easy to construct \eg{feeding in a null value}.
(3) \textit{Arid Code (33 Code Segments):}
Code segments classified as ``arid" according to Petrovic and Ivankovic's work~\cite{8449247}, referring to code with limited value for testing.
Arid code includes segments associated with logging, testing, non-functional properties, and axiomatic language features whose functionalities can be trivially tested \eg{getters and setters}.
(4) \textit{Overridden Standard APIs (21 code segments):}
This category includes cases where standard methods \eg{\texttt{toString()}, \texttt{equals()}, and \texttt{hashCode()}} are overridden.
(5) \textit{Data Structure Processing (15 code segments):}
Code segments that perform operations on data structures like arrays, lists, maps, and trees, such as traversal, sorting, searching, insertion, and deletion.
(6) \textit{Inherited and Overridden Functions (8 code segments):}
This category includes functions in class hierarchies that lack test coverage.
(7) \textit{Others (73 code segments):}
Code segments that do not belong to the above categories.
They are specific coding practices appearing only once or twice in our dataset.
The code segments in these subcategories are either trivially testable or require minimal effort to construct the necessary inputs, suggesting that their lack of coverage is likely due to oversight rather than inherent testing complexity.

\subsubsection{Category 3: Uncoverable Code Segments}
Finally, 107 out of the 716 (14.9\%) analyzed code segments cannot be covered due to (1) test suite anomalies (64 code segments), where errors and design flaws in test implementations or intentionally skipped tests prevent some tests from being executed; (2) pending implementations (26 code segments), where portions of the codebase are marked for future development with placeholders; and (3) dead code (17 code segments), where the code can never be reached \eg{unused variables, methods, classes, or unsatisfiable conditions.}

\begin{tcolorbox}[left=3pt,right=3pt,top=1pt,bottom=1pt]
	\textbf{Answer to RQ2:}
	\textit{%
		Over half of uncovered code segments (51.4\%) are impeded by coverage hurdles, predominantly due to external dependencies (98.6\%), especially network communications. Addressing these challenges requires advances in protocol-aware mocking and automated test environment setup.
	}
\end{tcolorbox}

\section{RQ3: Mocking in IoT Testing}
\label{sec:rq5}
\subsection{Approach}
Mocking is widely used in testing to isolate units from their dependencies.
Developers can create mock objects (also known as test doubles) to replace the original dependency implementations and fully control their behaviors~\cite{9794020}.
Studies have been conducted to characterize the usage of mock objects in different domains, such as Java programs~\cite{10.1145/3324884.3416539,zhu2023stubcoder,Xiao2024EmpiricalMocking,zhu2025mockassertions} or Android apps~\cite{9794020}.
In this research question, we aim to dissect the landscape of mock usage in IoT software testing.
To answer this research question, we (1) leveraged dynamic analysis to extract mock objects in the code, (2) conducted a closed coding analysis to identify the role of each mock object, and (3) conducted an open coding analysis to categorize the dependencies that are simulated by each mock object.

\textbf{Extracting mock objects from test cases.}
As pointed out by Zhu et al.~\cite{10.1145/3324884.3416539}, extracting mock objects via static analysis suffers from significant limitations in precision.
To avoid including extensive noise in our dataset, we leveraged a dynamic analysis approach to extract mock object instances from the test cases, following Zhu et al.~\cite{10.1145/3324884.3416539}.
Specifically, this approach identifies mock objects by inspecting the dependencies of each function call, including the types of its arguments and return value.
Mock objects created by mocking frameworks typically follow certain patterns in their type names.
For instance, in the commonly used \textsc{Java} mocking framework \textsc{Mockito}~\cite{mockito}, mock objects typically have type names that include \texttt{MockitoMock} or \texttt{EnhancerbyMockito}.
Similarly, for the \textsc{Python} \textsc{unittest} framework~\cite{unittest}, relevant mock object types are \texttt{unittest.mock.Mock} or \texttt{unittest.mock.MagicMock}.
\textsc{JavaScript} typically does not have sufficient runtime information to determine whether an object is a mock object or not, so our analysis seeks to identify invocations of functions that create mock objects, e.g.,~\texttt{sinon.stub()}~\cite{sinon}.
We can thereby identify mock objects by matching the type of the extracted function call dependencies with these framework-specific patterns.
We collected such patterns for 18 mock frameworks.
These frameworks were collected by identifying libraries that contain the keyword ``mock'' in our 317 projects and manually analyzing their documentation.
The details of the frameworks and their mock object type patterns are available in our research artifact~\cite{Github:IoTTestingAnalysis}.
This dynamic analysis resulted in 1,056 mock objects from all 37 projects collected in Section~\ref{sec:rq2}.
We used this dataset as the subject for our further analysis.

\textbf{Categorizing the mock objects.}
To understand the usage and purpose of mock objects, we conducted a manual analysis of the extracted mock objects. Specifically, we categorized them from two perspectives:
(1) \textit{Roles of the mock objects.}
Mock objects can play different roles in a test case.
In practice, there are five common types of mock objects~\cite{9794020}:
\begin{itemize}[leftmargin=*]
	\item \textit{Mock.} A mock is a replacement for the test dependencies such that the method invocations can be verified~\eg{using the API  \code{verify()} provided by \textsc{Mockito}}.
	\item \textit{Spy.} Spies are similar to mocks but wrap real objects, allowing the actual method to execute while still enabling behavior verification. In \textsc{Mockito}, spies can be created using the \code{spy()} method.
	\item \textit{Stub.} A stub is a mock object with hard-coded behaviors \ie{return values or exceptions} for its method invocations~\cite{zhu2023stubcoder}. Such hard-coded behaviors usually respond to specific scenarios in the tests.
	\item \textit{Fake.} A fake is a lightweight working implementation suitable for testing purposes only. For example, a fake database can use an array to store its data.
	\item \textit{Dummy.} A dummy is an object which neither the test nor the code under test interacts with. They are usually used as a placeholder for the parameter regardless of a specific test.
\end{itemize}
The roles of the mock objects can characterize how the mock objects are used.
(2) \textit{Types of dependencies simulated by the mock objects.} We also categorized mock objects based on the functionalities they simulate in the test environment.
This perspective characterizes what the mock objects are used for.
These two perspectives identify the challenges of mock usage in IoT testing and can guide future research on generating mocks for IoT-specific test cases.

We adopted an iterative procedure to manually analyze the mock objects that is similar to the procedure we describe in Section~\ref{ssec:rq3approach}.
Two authors independently analyzed the dataset to identify the roles and simulated dependencies of mock objects, forming and refining categorization criteria through discussion.
Throughout the process, they regularly discussed disagreements and refined the categorization criteria.
A third author was involved to resolve any conflicts between the initial two authors' classifications.
We achieved a Cohen’s Kappa score of 0.89 for the roles of the mock objects and 0.82 for the types of dependencies simulated by mock objects at the end, indicating a high level of agreement.

\subsection{Results}

\subsubsection{Roles of Mock Objects}

\begin{table}[t]
	\centering
	\caption{Roles of Mock Objects in IoT Software Testing}
	\label{tab:roleDistribution}
	\smaller
	\begin{tabularx}{\linewidth}{lRRRR}
		\toprule
		                         & \textbf{Mock} & \textbf{Stub} & \textbf{Spy} & \textbf{Fake} \\
		\midrule
		\textbf{\# Instances}    & 716           & 320           & 18           & 2             \\
		\textbf{Percentage (\%)} & 67.80         & 30.30         & 1.70         & 0.19          \\
		\bottomrule
	\end{tabularx}
\end{table}

Table~\ref{tab:roleDistribution} summarizes our analysis results of mock object roles in IoT software testing.
716 (67.80\%) out of the 1,056 analyzed mock objects play the role of Mock, and 30.30\% play the role of Stub.
These two categories account for 98.10\% of all the analyzed mock objects.
Conversely, Spy (1.70\%) and Fake objects (0.19\%) are comparatively rare.
We found no dummies in the dataset.
These results show that the vast majority of mock usages in IoT software test cases involve defining (both Mock and Stub) or verifying (Mock and Spy) specific dependency behaviors instead of simply acting as placeholders for the corresponding dependencies (Dummy and Fake).
As a result, it is non-trivial to generate mock objects to improve test generation for IoT software, since they need to encapsulate meaningful and context-specific dependency behaviors.

\begin{table}[t]
	\centering
	\caption{Mocked Dependency Types}\label{tab:mockApiTypes}
	\smaller
	\begin{tabularx}{\linewidth}{lR|lR}
		\toprule
		\textbf{Type}         & \textbf{\# Instances} & \textbf{Type}  & \textbf{\# Instances} \\
		\midrule
		Network Communication & 803 (76.04\%)         & User Interface & 15 (1.42\%)
		\\
		Logger                & 68 (6.44\%)           & Database       & 8  (0.76\%)           \\
		System Call           & 31 (2.94\%)           & Environment    & 8  (0.76\%)           \\
		Domain Logic          & 30 (2.84\%)           & CLI            & 7  (0.66\%)           \\
		Scheduling            & 29 (2.75\%)           & API Key        & 6  (0.57\%)           \\
		File                  & 15 (1.42\%)           & Others         & 36 (3.40\%)           \\
		\bottomrule
	\end{tabularx}
\end{table}

\subsubsection{Types of Dependencies
	Simulated by Mock Objects}
It is important to characterize the types of dependencies that are simulated by the mock objects to identify the frequently mocked APIs.
Results from this analysis can guide future research on leveraging mock objects to improve testing for IoT software.
Table \ref{tab:mockApiTypes} shows the categorization results for the types of simulated dependencies.
We identified 12 categories of different simulated dependencies as follows:

\textbf{Network Communication (803 mock objects).}
Mocks in this category simulate external communication interfaces, enabling tests to verify data exchange handling without real network calls.
This facilitates controlled testing of error handling, serialization, and protocol compliance.

\textbf{Logger (68 mock objects).}
Mocking logging dependencies allows tests to verify event recording behavior without relying on the real logging system.

\textbf{System Call (31 mock objects).}
Mocking system calls allows tests to simulate responses from the operating system, enabling the validation of project behavior in response to system-level events without real system interactions.

\textbf{Domain Logic (30 mock objects).}
In this context, domain logic mocks represent internal logic systems, allowing tests to handle these rules within the project without needing the actual implementations present.

\textbf{Scheduling (29 mock objects).}
This category contains mock objects used to test event scheduling and management.
It includes mocks that simulate scheduling logic, event queues, and event-driven behaviors.
\\ \textbf{File (15 mock objects).}
This category uses mocks to simulate file systems, enabling tests of file operations such as reading and writing without accessing real files.
This helps in verifying file manipulation behaviors under various conditions.

\textbf{User Interface (15 mock objects).}
User Interface mock objects simulate user interactions, which facilitates validating the response to user interactions without needing real users, leading to a higher degree of test automation.

\textbf{Database (8 mock objects).}
Mock objects simulating database interactions enable testing data handling logic—such as queries and transactions—independently from the actual database, ensuring correct implementation.

\textbf{Environment (8 mock objects).}
Environment mock objects simulate different runtime environments, testing the project's adaptability and correctness under various configurations without altering the real execution environment.

\textbf{Command Line Interface (CLI, 7 mock objects).}
Mock objects for CLI components simulate the behavior of the command line and terminal.
Similar to mock objects for UI, CLI mock objects can verify the processing logic of CLI arguments and commands without actual user input.

\textbf{API Key (6 mock objects).}
Mocking API key management systems allows for the testing of key storage, retrieval, and usage, ensuring secure and appropriate use of API keys without real external service interactions.

\textbf{Others (36 mock objects).}
This category groups mock objects that simulate external dependencies not easily classified into existing categories.
These mocks represent unique or infrequent dependencies with limited presence in typical IoT testing scenarios.

\begin{tcolorbox}[left=3pt,right=3pt,top=1pt,bottom=1pt]
	\textbf{Answer to RQ3:}
	\textit{
		Mock objects are primarily used to test communication APIs in IoT software, and are also frequently applied to simulate system calls, files, databases, and environment APIs, thus addressing key code coverage challenges (RQ2). Mocks are essential for overcoming IoT testing barriers, but generating effective mocks remains difficult due to the requirement for realistic and context-specific behaviors.
	}
\end{tcolorbox}

\section{Implications}
In this section, we discuss the implications that can be drawn from the results of our research questions.

\textbf{Implication 1.}
\textit{\bfseries The prevalent use of mock objects aligns with the core testing challenges in IoT software, with both showing a striking focus on network communication scenarios. Leveraging mock objects presents a promising approach to address these challenges.}
As shown in Table~\ref{tab:mockApiTypes}, among all the mocked dependencies, 76.04\% simulate network APIs.
This number is significantly higher than that in other types of software.
For example, only 8.7\% of the mock objects in Android apps target network APIs~\cite{9794020}.
This striking difference reflects the unique demands of IoT systems, which heavily rely on network communications through various protocols such as Zigbee and MQTT.
In the meantime, these results align closely with the categorization results in Table~\ref{tab:cate1} (RQ2):
363 out of 368 code segments in Category 1 involve external dependencies, while
199 segments (54.07\%) are due to untested communication scenarios across diverse protocols.
Our findings reveal a strong overlap between the challenges in covering IoT code and the types of dependencies simulated by mock objects.
This overlap highlights network communication as the primary testing challenge for IoT software and suggests mocking as a de facto solution to address this challenge.
It may seem contradictory that communication-related code remains largely untested while such mock objects are widely used. This is because the results are collected across different projects. While some well-maintained projects effectively mock complex network dependencies, many others still face difficulties.
This observation also leads to our next implication.

\textbf{Implication 2.
} \textit{\bfseries Despite the challenges of correctly mocking network communication in IoT software, the abundance of existing mocks in open-source projects enables mock migration as a potential solution.}
As demonstrated in RQ3, mock objects in IoT tests need to faithfully encapsulate the specific dependency behaviors rather than act as simple placeholders.
Defining such mocks to replicate communication scenarios involving IoT devices remains a significant challenge.
Existing mock generation techniques, which often depend on human-written test oracles~\cite{zhu2023stubcoder} or runtime profiling~\cite{tiwari2024mimicking}, are not applicable since it is infeasible to predict or profile the behaviors of all IoT devices.
Based on our empirical study results, we envision migrating mock objects across software as a solution to address this problem since: (1) IoT software systems often interact with a common set of IoT devices, (2) the behaviors of these devices are likely to remain consistent across different software contexts, and (3) as demonstrated by RQ3, a large number of mock objects are already available in open-source projects, providing a rich resource for migration.
For example, Shelly is an IoT device producer providing a variety of IoT devices for smart homes~\cite{shelly}.
Both Home Assistant~\cite{Github:home-assistant/core} and OpenHAB~\cite{Github:openHAB} smart home platforms support integrations of Shelly devices.
However, the Shelly component in Home Assistant contains substantial test cases while the component in openHAB remains untested.
Lines 1-3 in Listing~\ref{lst:migration} demonstrate how Home Assistant uses a mock function to simulate setting the state of a Shelly light device.
This function accepts structured parameters (e.g., \texttt{red}, \texttt{green}, \texttt{blue}, \texttt{brightness}) that conform to the Shelly protocol, enabling Home Assistant to test its Shelly integration logic without relying on real devices.
In comparison, openHAB’s \texttt{ShellyLightHandler} performs similar operations to control device states (Lines 6-7).
However, this logic remains untested due to the absence of a mocking infrastructure.
Notably, both platforms interact with Shelly devices using the same protocol, and their commands share the same attributes (e.g., colors, brightness) and valid ranges.
This alignment presents a clear opportunity: test code from Home Assistant can be adapted to simulate equivalent device states in openHAB.
By reusing mock inputs and verifying method calls, it becomes feasible to migrate test cases and improve test coverage for \texttt{ShellyLightHandler}.

\begin{figure}[t]
	\fontsize{7}{7}
	\begin{lstlisting}[language=Python, caption={Example of Test Migration}, label=lst:migration ,gobble=8, numbers=left,breaklines=true,
    breakatwhitespace=false,
    columns=flexible]
        # Mock function in Home Assistant to simulate setting light state
        def mock_light_set_state(... red=45, green=55, brightness=50...):
          return {... "red": red, "green": green,  "brightness": brightness...}

        # Real implementation in openHAB's ShellyLightHandler to control light
        col.setRed(setColor(lightId, "red", command, 255));
        col.setGreen(setColor(lightId, "green", command, 255));
        ...
        \end{lstlisting}
\end{figure}

\begin{table}[t]
\centering
\caption{Coverage Hurdles and Potential Solutions}
\label{tab:solutions}
\begin{tabular}{p{0.4\linewidth} p{0.5\linewidth}}
\toprule
\textbf{Coverage Hurdle} & \textbf{Potential Solution} \\
\midrule
Untested Communication Scenarios & Systematic mocking support and reuse of existing device/service mocks \\
\midrule
Misaligned File System Dependency & Mocking file resources and filesystem states \\
\midrule
Insufficient Database Setup & Mocking and seeding representative database states \\
\midrule
Improper System Configuration & Systematic configuration-space exploration (e.g., iGen) \\
\midrule
Untested IoT-Event Handlers & Automated event-sequence generation (e.g., IoT-TEG) \\
\bottomrule
\end{tabular}
\end{table}

\textbf{Potential Solutions to Identified Coverage Hurdles.}
To make our findings more actionable, Table~\ref{tab:solutions} summarizes alternative solution directions for each identified hurdle for improving the test coverage.

For some hurdles, existing techniques already provide promising starting points.
For \textbf{Improper System Configuration}, systematic exploration of configuration spaces can help exercise configuration-dependent branches, for example through configuration interaction inference and test generation techniques such as iGen~\cite{10.1145/2950290.2950311}.
For \textbf{Untested IoT-Event Handlers}, automated generation and mutation of event sequences can stimulate event-driven behaviors, as demonstrated by IoT event generation frameworks such as IoT-TEG~\cite{velez2022iotteg}.

For the remaining hurdles, including \textbf{Untested Communication Scenarios}, \textbf{Misaligned File System Dependency}, and \textbf{Insufficient Database Setup}, the core difficulty lies in reproducing realistic external dependencies during testing.
While existing approaches such as fuzzing or automated input generation can partially address these issues, they often fail to capture the protocol semantics, data formats, and stateful behaviors required by IoT software.
As discussed in Implications~1 and~2, developers therefore rely on mock objects to emulate external services, devices, files, and databases.
However, manually constructed mocks are often incomplete and inconsistent across projects, leaving many interaction paths uncovered.
Based on our findings, we suggest that improving the fidelity and reuse of mocks is a more practical and impactful solution direction.
In particular, the abundance of existing mock objects in open-source IoT projects enables systematic mock improvement and migration across software that interacts with the same devices or protocols, providing an opportunity to reduce uncovered code in practice.

\textbf{Mocking Scope and Limitations.}
Mock-based testing is widely adopted in IoT software projects to isolate external dependencies, such as MQTT or HTTP endpoints, environment variables, databases, and file systems.
Such mocks improve controllability and reproducibility in software-level tests and enable developers to exercise code paths that would otherwise depend on unavailable or non-deterministic conditions.

However, mock-based isolation tests differ fundamentally from execution on real devices or hardware-in-the-loop setups.
When IoT software is deployed on physical devices, factors such as timing constraints, network instability, and hardware-specific behaviors may not be accurately reproduced by simulated components.
As a result, while mocks are essential for scalable software testing, final validation of IoT behavior still requires execution on real devices or realistic testbeds.

Therefore, our findings should be interpreted as reflecting testing practices in software repositories, where mocks primarily support reproducible development testing, while device-level execution may remain necessary for confirming end-to-end IoT behavior in practice.

\section{Threats to Validity}
\label{sec:threats-to-validity}
\textbf{External Validity.}
Our study focused on open-source IoT projects and the test cases within those projects.
Device-layer testing practices such as Model-in-the-Loop testing may be conducted using external environments and thus may not be reflected in repository contents.
As a result, our findings primarily reflect the testing practices that are observable in open-source IoT software projects, while still providing useful empirical evidence on how developers test and validate IoT software in practice.
Our analyzed testing and mocking frameworks may not cover all the frameworks used by developers.
As a result, our findings may not generalize to closed-source IoT software or those using different frameworks for testing or mocking.
However, the IoT projects we collected span various domains and sizes to ensure that our findings are as representative as possible of the open-source IoT ecosystem.
Additionally, our analysis is conducted at the repository level and does not explicitly separate IoT-specific code from general-purpose utility or application code within the same project.
In practice, IoT repositories often combine protocol-handling and device-facing logic with conventional components (e.g., configuration, storage, or web services).
Therefore, some uncovered-code categories we observed may also arise in non-IoT software, although they manifest in IoT projects under stronger environmental and dependency constraints.
Nevertheless, RQ2 results highlighted major IoT-relevant challenges, e.g., ``Untested Communication Scenarios'' and ``Untested IoT-Event Handlers'', suggesting that the non-IoT-specific code in the projects did not significantly impact our key findings.
Our study focuses on functional testing artifacts observable in open-source IoT repositories.
Testing of non-functional properties (e.g., performance, security, or real-time constraints) and hardware-dependent validation is not systematically captured in such repositories.
As a result, such testing cannot be analyzed in a uniform and reproducible manner across projects using repository-based evidence alone.
Our findings should thus be interpreted as characterizing functional testing practices in open-source IoT software components across different IoT layers.

\textbf{Internal Validity.}
The categorization of uncovered code segments and mock objects involved manual open-coding and closed-coding processes.
While these methods are grounded in rigorous qualitative research principles, they are susceptible to subjective bias.
To reduce the impact of subjective bias, two authors independently coded a subset of the data, and discrepancies were discussed until a consensus was reached.
Moreover, our approach included sampling when analyzing the uncovered code segments, which might have led to the omission of relevant data that could influence the results.
We also ensured that our confidence level and sampling strategy were systematically applied to obtain a statistically representative sample size and avoid biases related to the selection of code segments.

\textbf{Construct Validity.}
In this study, test effectiveness is defined at the source-code level and measured using statement coverage, branch coverage, and mutation score.
These metrics reflect the extent to which tests exercise implementation logic and reveal faults, but do not capture system-level behavior or non-functional properties.
We follow prior empirical software testing studies that adopt coverage and mutation analysis as widely accepted metrics for test-suite effectiveness~\cite{10.1145/3635713}.
These metrics enable objective comparison across projects and provide a basis for assessing the effectiveness of IoT test suites in practice.
To measure test effectiveness, we used established coverage and mutation testing tools.
Specifically, statement and branch coverage were collected using \textsc{Coverage.py} for Python,
\textsc{JaCoCo} for Java, and \textsc{nyc} for JavaScript.
Mutation scores were computed using \textsc{mutmut} (Python), \textsc{PIT} (Java), and
\textsc{Stryker} (JavaScript).
Static analysis and test artifact inspection were conducted using custom scripts built on top of
language-specific abstract syntax trees and repository metadata.
To mitigate potential inaccuracies in our tool implementation, we conducted tests of our implementation and manually checked the results.
When using existing coverage or mutation testing tools, we strictly followed their instructions.
However, it is possible that running the test cases in different environments could potentially lead to different code coverage results.
To mitigate this threat, we strictly followed the instructions provided by the analyzed IoT software projects, aiming to replicate the coverage that can be achieved by the test cases.
We also provide Docker images that can reproduce our evaluation results~\cite{Github:IoTTestingAnalysis}.

\section{Related Work}
\label{sec:related-work}

\textbf{\revised{IoT Software.}}
\revised{
\emph{Testing Techniques and Frameworks.}
Research on IoT testing has explored a variety of testing techniques and tools, including several widely studied model-based testing~\cite{10.1007/978-3-319-47169-3_55}, smart home application testing~\cite{manandhar2020towards,mandal2023helion}, and comparisons of IoT testing tools~\cite{8411738}.
More recent work has proposed IoT-specific testing frameworks, such as event-driven test generation for realistic IoT behaviors~\cite{velez2022iotteg}, model-based generation of executable end-to-end IoT test scripts~\cite{10.1007/s11219-021-09565-y}, combinatorial testing strategies for complex IoT systems~\cite{10.1007/s10664-021-10017-1}, automated interoperability and integration testing platforms~\cite{bures2021patriot}, scalable online conformance testing for IoT applications~\cite{hwang2020autoconiot}, a testing framework for remote IoT test execution~\cite{8281514}, protocol-level testing under varying network conditions and traffic patterns~\cite{7172291}, smart testing frameworks with resource-demand prediction in dynamic IoT environments~\cite{8605788}, and model-driven Thing-in-the-Loop approaches for IoT verification and validation~\cite{10.1145/3137003.3137007}.
Protocol- and network-level robustness testing has also been studied using fuzzing-based approaches~\cite{shu2022iotinfer,paduraru2021riveriot}.
Complementary to pre-deployment testing, runtime verification approaches monitor IoT executions during operation, and runtime testing has also been systematically reviewed~\cite{9411895,10.1007/s11219-021-09558-x}.}

\revised{
\emph{Failures and Diagnosis.}
Several studies have investigated challenges and bugs in IoT software, with a primary focus on security and reliability issues.
Prior work has examined vulnerabilities arising from interactions with the physical world~\cite{ding2018safety}, firmware-level defects~\cite{7932855}, protocol implementations~\cite{7958578}, and IoT applications~\cite{10.1145/3395363.3397347}.
Beyond security-oriented analyses, Makhshari et al. conducted a large-scale empirical study of IoT bug reports, analyzing 5,565 issues from 91 repositories to construct a taxonomy of failures, root causes, and fault locations~\cite{DBLP:conf/icse/Makhshari021}.
More recent work has also investigated debugging, troubleshooting, and fault diagnosis in deployed IoT systems, including live debugging support for IoT applications~\cite{10.48550}, record-and-replay techniques for diagnosing COTS IoT devices~\cite{fang2020iotreplay}, systematic fault handling mechanisms~\cite{10.1145/3532194}, and distributed fault diagnosis frameworks~\cite{9745085}.}

\revised{
\emph{Reviews and Surveys.}
Several systematic literature reviews have surveyed IoT systems testing and related quality concerns, summarizing techniques, challenges, and research trends~\cite{10433067,10.1145/3625094,10759538,10710185}.
In addition to technical approaches, survey studies have gathered practitioner perspectives on IoT testing challenges.
Minani et al.~\cite{10168828} surveyed 49 practitioners and identified issues such as heterogeneity, lack of standards, and limited automation, while Rodriguez-Cardenas et al.~\cite{10988986} surveyed 80 developers and reported issues related to device compatibility, firmware updates, and documentation.}

\revised{
Despite this growing body of work, existing studies largely focus on proposing testing techniques, tools, or studying practitioner perceptions.
They do not empirically characterize \emph{test effectiveness}, \emph{test coverage}, or \emph{mocking practices} based on executing tests in real IoT repositories.
In contrast, our work provides a measurement-based analysis by executing and analyzing test suites from 37 runnable IoT projects, offering empirical evidence on how IoT software is tested in practice.}

\textbf{Empirical Studies On Testing.}
Empirical studies have been conducted to characterize testing practices in different domains.
For example, researchers have studied Android apps to explore automation~\cite{9286051} and GUI testing~\cite{Coppola2019ScriptedGUITesting}, and Java projects to correlate test coverage with bugs~\cite{8031982} and software metrics~\cite{7091313}, such as the number of developers involved. While our data collection approach may share similarities with these previous efforts, our focus distinctly lies on identifying and understanding testing challenges within IoT software.
Kochhar \etal{}~\cite{7102609} analyzed test case existence and computed code coverage across 600 Android apps, using surveys to understand the challenges faced by developers.
Unlike the study that relied on surveys to gauge testing challenges, we directly analyzed uncovered code segments to reveal testing difficulties.

These studies have explored testing practices across diverse software domains.
However, testing in IoT remains underexplored.
Our research specifically focuses on IoT software developed in Python, Java, and JavaScript, which distinguishes it from the Android applications targeted in previous studies.

\textbf{Empirical Studies On Mocking.}
Several studies have analyzed the use of mocking frameworks in software testing, highlighting the prevalence and implementation practices across various projects.
For instance, Xiao \etal{}~\cite{Xiao2024EmpiricalMocking} analyzed 246 Apache Java projects to understand how mocking frameworks are utilized and the common usage patterns within these projects.
Fazzini \etal{}~\cite{9794020} employed static analysis to extract and categorize mock objects from 1,000 Android applications, providing a broad overview of mocking practices in mobile app development.
In contrast, our approach leverages dynamic analysis to extract mock objects, which offers higher precision.
Zhu \etal{}~\cite{10.1145/3324884.3416539} conducted an extensive study on mocking decisions with 10,000 test cases across four open-source projects and proposed \textsc{MockSniffer} to recommend mocking decisions.
More recently, Zhu \etal{}~\cite{zhu2025mockassertions} further conducted an empirical study on the usage of mock assertions to reveal their practical usage and fault detection capabilities.

These studies have primarily focused on static analysis and generic mocking recommendations, with some providing categorizations of mock usage.
In comparison, our research goes beyond this by developing a taxonomy specifically tailored to the unique needs of IoT software.
This taxonomy categorizes mock objects based on their functionalities, with particular emphasis on how they address the challenges of simulating external dependencies unique to IoT environments.

Our study addresses gaps in the existing IoT literature by providing a focused empirical study of testing practices within IoT software, particularly in understanding the utilization of mocks and the challenges associated with testing.

\section{Conclusion}
\label{sec:conclusion}
In this paper, we presented a large-scale empirical study of testing practices in open-source IoT software projects.
Our investigation revealed that, although IoT software tends to include more tests than general-purpose software, its coverage and effectiveness remain limited.
We identified significant challenges in IoT software testing, particularly in handling complex dependencies and interactions such as network communications.
Furthermore, our study highlighted the substantial role of mock objects in IoT software testing and revealed a clear alignment between mock usage and the challenges faced by developers.
Future research can build on these insights to enhance IoT software testing by improving the utilization of mock objects, developing techniques that can better simulate complex system interactions, and creating tools that facilitate more effective and automated testing for IoT software.
We envisioned mock object migration as a promising direction for future research, leveraging the large number of mock objects available in well-maintained open-source projects to generate tests with faithfully simulated dependencies.

\section*{Acknowledgements}
This work was supported by Fonds de recherche du Québec(Grant No.2024-NOVA-346499)~\cite{nova}.

\balance
\bibliographystyle{IEEEtran}
\bibliography{References/Papers,References/Links}

\end{document}